\documentclass{aa}  
\usepackage{graphicx}
\usepackage{txfonts}
\usepackage{xcolor}
\usepackage{hyperref}
\hypersetup{colorlinks, linkcolor=blue, citecolor=blue, urlcolor=blue}

\newcommand{\te}{t_{\rm E}}
\newcommand{\thetae}{\theta_{\rm E}}

\newcommand{\pie}{\pi_{\rm E}}

\newcommand{\dl}{D_{\rm L}}
\newcommand{\ds}{D_{\rm S}}
\def\e{{\rm E}}

\definecolor{brown}{rgb}{0.59, 0.29, 0.0}
\definecolor{darkgreen}{rgb}{0.0, 0.42, 0.24}
\definecolor{darkblue}{rgb}{0.01, 0.31, 0.59}
\definecolor{darkblue}{rgb}{0.0, 0.25, 0.42}
\definecolor{blue}{rgb}{0.0,0.0,1.0}
\definecolor{green}{rgb}{0.0,1.0,0.0}

\begin{document}

\title{KMT-2021-BLG-2609Lb and KMT-2022-BLG-0303Lb: Microlensing planets 
identified through signals produced by  major-image perturbations}
\titlerunning{Microlensing planets identified through signals produced by  major-image perturbations}

\author{
     Cheongho~Han\inst{\ref{inst1}} 
\and Michael~D.~Albrow\inst{\ref{inst2}}   
\and Chung-Uk~Lee\inst{\ref{inst3},\ref{inst12}}
\and Sun-Ju~Chung\inst{\ref{inst3}}      
\and Andrew~Gould\inst{\ref{inst4},\ref{inst5}}      
\and Kyu-Ha~Hwang\inst{\ref{inst3}} 
\and Youn~Kil~Jung\inst{\ref{inst3}} 
\and Chung-Uk~Lee\inst{\ref{inst3}} 
\and Yoon-Hyun~Ryu\inst{\ref{inst3}} 
\and Yossi~Shvartzvald\inst{\ref{inst6}}   
\and In-Gu~Shin\inst{\ref{inst7}} 
\and Jennifer~C.~Yee\inst{\ref{inst7}}   
\and Hongjing~Yang\inst{\ref{inst10}}     
\and Weicheng~Zang\inst{\ref{inst7},\ref{inst10}}     
\and Sang-Mok~Cha\inst{\ref{inst3},\ref{inst11}} 
\and Doeon~Kim\inst{\ref{inst1}}
\and Dong-Jin~Kim\inst{\ref{inst3}} 
\and Seung-Lee~Kim\inst{\ref{inst3}} 
\and Dong-Joo~Lee\inst{\ref{inst3}} 
\and Yongseok~Lee\inst{\ref{inst3},\ref{inst11}} 
\and Byeong-Gon~Park\inst{\ref{inst3}} 
\and Richard~W.~Pogge\inst{\ref{inst5}}
\\
(The KMTNet Collaboration)
}

\institute{
      Department of Physics, Chungbuk National University, Cheongju 28644, Republic of Korea\label{inst1}                                                          
\and  University of Canterbury, Department of Physics and Astronomy, Private Bag 4800, Christchurch 8020, New Zealand\label{inst2}                                 
\and  Korea Astronomy and Space Science Institute, Daejon 34055, Republic of Korea\label{inst3}                                                                    
\and  Max Planck Institute for Astronomy, K\"onigstuhl 17, D-69117 Heidelberg, Germany\label{inst4}                                                                
\and  Department of Astronomy, The Ohio State University, 140 W. 18th Ave., Columbus, OH 43210, USA\label{inst5}                                                   
\and  Department of Particle Physics and Astrophysics, Weizmann Institute of Science, Rehovot 76100, Israel\label{inst6}                                           
\and  Center for Astrophysics $|$ Harvard \& Smithsonian 60 Garden St., Cambridge, MA 02138, USA\label{inst7}                                                      
\and  Department of Astronomy and Tsinghua Centre for Astrophysics, Tsinghua University, Beijing 100084, China\label{inst10}                                       
\and  School of Space Research, Kyung Hee University, Yongin, Kyeonggi 17104, Republic of Korea\label{inst11}                                                      
\and  Corresponding author\label{inst12}                                                                                                                           
}                                                                                                                                                                                                                 
\date{Received ; accepted}                                                                                                                                                                                      
                                                                                                                                                                            

\abstract
{}
{
We investigate microlensing data collected by the Korea Microlensing Telescope Network 
(KMTNet) survey during the 2021 and 2022 seasons to identify planetary lensing events 
displaying a consistent anomalous pattern. Our investigation reveals that the light curves 
of two lensing events, KMT-2021-BLG-2609 and KMT-2022-BLG-0303, exhibit a similar anomaly, 
in which short-term positive deviations appear on the sides of the low-magnification lensing 
light curves.
}
{
To unravel the nature of these anomalies, we meticulously analyze each of the lensing events. 
Our investigations reveal that these anomalies stem from a shared channel, wherein the source 
passed near the planetary caustic induced by a planet with projected separations from the host 
star exceeding the Einstein radius.  We find that interpreting the anomaly of KMT-2021-BLG-2609 
is complicated by the "inner--outer" degeneracy, whereas for KMT-2022-BLG-0303, there is no such 
issue despite similar lens-system configurations.  In addition to this degeneracy, interpreting 
the anomaly in KMT-2021-BLG-2609 involves an additional degeneracy between a pair of solutions, 
in which the source partially envelops the caustic and the other three solutions in which the 
source fully envelopes the caustic. As in an earlier case of this so-called von Schlieffen--Cannae 
degeneracy, the former solutions have substantially higher mass ratio.
}
{
Through Bayesian analyses conducted based on the measured lensing observables of the event 
time scale and angular Einstein radius, the host of KMT-2021-BLG-2609L is determined to be 
a low-mass star with a mass $\sim 0.2~M_\odot$ in terms of a median posterior value, while 
the planet's mass ranges from approximately 0.032 to 0.112 times that of Jupiter, depending 
on the solutions.  For the planetary system KMT-2022-BLG-0303L, it features a planet with a 
mass of approximately $0.51~M_{\rm J}$ and a host star with a mass of about $0.37~M_\odot$.  
In both cases, the lenses are most likely situated in the bulge.
}
{}

\keywords{planets and satellites: detection -- gravitational lensing: micro}

\maketitle

\begin{figure*}[t]
\centering
\sidecaption
\includegraphics[width=12.0cm]{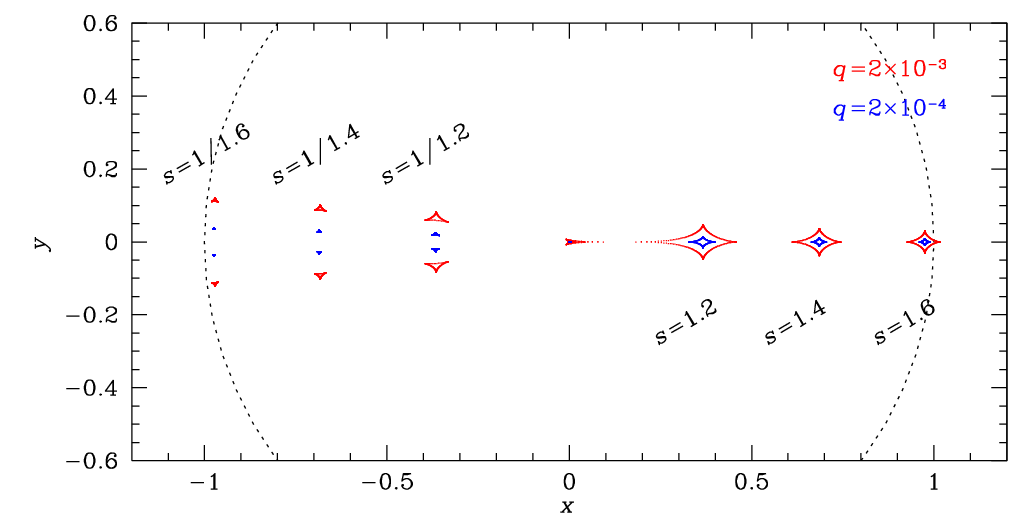}
\caption{
Variation of lensing caustics induced by planets. We illustrate two sets of caustics with 
planet-to-host mass ratios $q = 2 \times 10^{-3}$ (depicted in red) and $q = 2\times 10^{-4}$ 
(depicted in blue). Within each set, a series of caustics demonstrates variations depending 
on the planetary separation $s$.  The coordinates are centered at the position of the planet 
host and the lengths are scaled to the Einstein radius.  The dotted circle centered at the 
origin represents the Einstein ring.  
}
\label{fig:one}
\end{figure*}

\section{Introduction} \label{sec:one}

Since the emergence of high-cadence surveys in the 2010s, there has been a significant increase 
in the detection of microlensing planets \citep{Gould2022b, Jung2022}.  According to the "NASA 
Exoplanet Archive"\footnote{\tt https://exoplanetarchive.ipac.caltech.edu/}, the number of known 
microlensing planets has risen to 223, making microlensing the third most productive method for 
planet discovery. Microlensing signals from planets occur when a source star passes close to a 
lensing caustic \citep{Mao1991, Gould1992}.  In gravitational microlensing, caustics are specific 
regions at which the magnification of a background point source becomes infinitely large due to the 
lensing effect. The characteristics of these caustics, including their position, size, and shape, 
depend on the separation of the planet from its host and their mass ratio \citep{Chung2005, Han2006}. 
Additionally, the various trajectories of source stars relative to the lens system contribute to 
the diverse range of patterns observed in microlensing signals.

The increasing number of microlensing planets, along with the wide variety of signal forms, has
led to a trend of jointly announcing discoveries that show similar anomaly patterns in their light 
curves.  \citet{Han2017} and \citet{Poleski2017} provided notable examples of planetary signals 
arising from a recurring pathway in their analyses of the microlensing events OGLE-2016-BLG-0263 
and MOA-2012-BLG-006, respectively. \citet{Han2024a} identified planetary signals characterized 
by brief dips flanked by shallow rises on either side in their analyses of three events: 
MOA-2022-BLG-563, KMT-2023-BLG-0469, and KMT-2023-BLG-0735. Weak short-term planetary signals 
generated without caustic crossings were exemplified by \citet{Han2023} and \citet{Han2021} for 
the microlensing events KMT-2022-BLG-0475, KMT-2022-BLG-1480, KMT-2018-BLG-1976, KMT-2018-BLG-1996, 
and OGLE-2019-BLG-0954. \citet{Han2024b} announced the discoveries of three planets -- KMT-2023-BLG-0416, 
KMT-2023-BLG-1454, and KMT-2023-BLG-1642 -- identified from partially covered signals.

Recently, \citet{Han2024c} announced the discovery of four microlensing planets -- KMT-2020-BLG-0757Lb, 
KMT-2022-BLG-0732Lb, KMT-2022-BLG-1787Lb, and KMT-2022-BLG-1852Lb -- by analyzing signals observed 
on the wings of the lensing light curves.  These signals originated from "minor-image" perturbations, 
occurring when source stars pass through peripheral caustics induced by "close" planets. In a 
separate study, \citet{Jung2021} identified planetary signals in the lensing events OGLE-2018-BLG-0567 
and OGLE-2018-BLG-0962, for which the signals with positive deviations resulting from major-image 
perturbations appear on the sides of the light curves. Planets through signals of a similar type were 
reported from the analyses of the lensing events OGLE-2017-BLG-1777, OGLE-2017-BLG-0543 by \citet{Ryu2024} 
and OGLE-2017-BLG-0448 by \citet{Zhai2024}.  These perturbations occur when source stars traverse over 
peripheral caustics induced by "wide" planets. In the subsequent section, we explain the microlensing 
terminologies "close", "wide", "major image," and "minor image".  This systematic classification of 
planetary signals facilitates the recognition of similar patterns within the light curves of upcoming 
lensing events.

In this study, we introduce two additional microlensing planets identified via signals produced 
through major-image perturbations: KMT-2021-BLG-2609Lb and KMT-2022-BLG-0303Lb.  We discuss shared 
characteristics of the signals and explore the origins of these signals. Additionally, we offer 
detailed explanations of the typical types of degeneracies frequently encountered when interpreting 
planetary signals arising through this channel.

\section{Signals through major-image perturbations} \label{sec:two}

In the case where a single source undergoes microlensing by a single mass (referred to as a 1L1S
event), the image of the source is split into two distinct images. One of these images, known as
the "major image", appears brighter and is located outside the Einstein ring, while the other,
termed the "minor image", appears fainter and is positioned inside the ring. Within the context of
the lens plane, a planetary signal of a lens system comprising two masses (referred to as a 2L1S
system) of a planet and its host emerges when the planet perturbs either of the source images.
When the planet perturbs the major image, it induces further magnification of the image, resulting
in positive anomalies in the lensing light curve. Conversely, when the  minor image is perturbed by
the planet, it experiences demagnification, leading to negative deviations.  Given the positions of 
the two images, major-image perturbations occur when a planet has a separation greater than the 
Einstein radius ($\thetae$), while minor-image perturbations occur when the planet's separation is 
less than the Einstein radius. In microlensing studies, lengths are typically scaled to $\thetae$, 
and the projected planet-host separation normalized to $\thetae$ is represented as $s$ (normalized 
separation).  Consequently, the terms "close" and "wide" denote planets with $s<1$ and $s>1$, 
respectively.

The region of planetary perturbation in the lens plane corresponds to the position of
caustics in the source plane. A planet induces two sets of caustics: one centered around the
position of the planet's host star ("central caustic"), and the other positioned away from
the host ("planetary caustic") at approximately
\begin{equation}
{\bf u}_{\rm c} = {\bf s} - {1\over {\bf s}}. 
\label{eq1}
\end{equation}
Here ${\bf s}$ represents the position vector of the planet with respect to the host. If 
the mass ratio between a planet and its host star ($q$) is very low and the normalized 
separation is not precisely unity, then the central caustic induced by a wide planet with $s$ 
and that induced by a close planet with $1/s$ display notable similarities \citep{Dominik1999, 
An2005}.  However, the pair of planetary caustics induced by the close and wide planets differ 
from each other in number, shape, and location. A close planet results in a pair of three-fold 
caustics on the opposite side of the planet  with respect to the host, while a wide planet produces 
a single four-fold caustic on the planet side.  For both types of planets, the size of the planetary 
caustics decreases in proportion to the square root of the mass ratio between the planet and host 
\citep{Han2006}.  For a given planetary lens, the planetary caustic is substantially bigger than 
the central caustic. Figure~\ref{fig:one} illustrates how the caustics vary depending on the 
planetary separation and mass ratio. Because of the positions of caustics, planetary signals 
resulting from central caustics appear near the peak of the light curve for high-magnification 
events \citep{Griest1998}, whereas signals caused by planetary caustics tend to emerge on the 
side of the light curve for events with low lensing magnifications.

\section{Observations and data} \label{sec:three}

The planetary lensing events analyzed in this study were identified by examining data collected 
from the Korea Microlensing Telescope Network \citep[KMTNet:][]{Kim2016} survey conducted during 
the 2021 and 2022 seasons. The KMTNet group operates a lensing experiment employing a network of 
three telescopes strategically located across the Southern Hemisphere: one at the Cerro Tololo 
Inter-American Observatory in Chile (KMTC), another at the South African Astronomical Observatory 
in South Africa (KMTS), and the third at the Siding Spring Observatory in Australia (KMTA). These 
telescopes are identical, each featuring a 1.6-meter aperture and equipped with a camera providing 
a field of view of 4 square degrees.

The microlensing survey involves monitoring star brightness in the Galactic bulge direction to 
detect gravitational lensing-induced light variations.  The majority of star images were captured 
in the $I$ band, with approximately one-tenth of images taken in the $V$ band specifically for 
source color measurements.  Data reduction and photometry employed the pipeline developed by 
\citet{Albrow2009}, which incorporates the difference imaging analysis technique \citep{Tomaney1996, 
Alard1998, Wozniak2000}.  To ensure optimal data quality for our analysis, we performed a re-reduction 
of the KMTNet data using the photometry software developed by \citet{Yang2024}.  Error bars were 
adjusted not only to ensure consistency between the scatter of data and errors but also to establish 
$\chi^2$ per degree of freedom of unity for each data set.  This process was carried out following 
the routine outlined in \citet{Yee2012}.

\begin{table*}[h]
\caption{2L1S lensing parameters of KMT-2021-BLG-2609\label{table:one}}
\begin{tabular}{l|lll|lll}
\hline\hline
\multicolumn{1}{c|}{Parameter}         &
\multicolumn{3}{c|}{Small $q$}         &
\multicolumn{2}{c}{Large $q$}          \\
\multicolumn{1}{c|}{}                  &
\multicolumn{1}{c}{intermediate}       &
\multicolumn{1}{c}{inner}              &
\multicolumn{1}{c|}{outer}             &
\multicolumn{1}{c}{inner}              &
\multicolumn{1}{c}{outer}              \\
\hline
 $\chi^2$                    &  $1703.4            $    &  $1704.2            $     &  $1703.8            $  &  $1705.0            $   &  $1704.5            $   \\   
 $t_0$ (HJD$^\prime$)        &  $9498.037 \pm 0.218$    &  $9498.004 \pm 0.212$     &  $9497.873 \pm 0.210$  &  $9497.881 \pm 0.215$   &  $9498.099 \pm 0.206$   \\   
 $u_0$                       &  $0.813 \pm 0.093   $    &  $0.757 \pm 0.139   $     &  $0.750 \pm 0.128   $  &  $0.637 \pm 0.116   $   &  $0.711 \pm 0.077   $   \\   
 $\te$ (days)                &  $15.61 \pm 1.20    $    &  $16.18 \pm 1.41    $     &  $16.32 \pm 1.51    $  &  $17.95 \pm 1.57    $   &  $16.82 \pm 1.13    $   \\   
 $s$                         &  $1.519 \pm 0.067   $    &  $1.498 \pm 0.105   $     &  $1.467 \pm 0.090   $  &  $1.424 \pm 0.085   $   &  $1.422 \pm 0.050   $   \\   
 $q$ (10$^{-3}$)             &  $0.152 \pm 0.038   $    &  $0.221 \pm 0.045   $     &  $0.191 \pm 0.049   $  &  $0.529 \pm 0.264   $   &  $0.579 \pm 0.190   $   \\   
 $\alpha$ (rad)              &  $5.065 \pm 0.020   $    &  $5.070 \pm 0.021   $     &  $5.067 \pm 0.021   $  &  $5.083 \pm 0.023   $   &  $5.092 \pm 0.019   $   \\   
 $\rho$ (10$^{-3}$)          &  $33.90 \pm 3.99    $    &  $34.79 \pm 5.67    $     &  $34.18 \pm 4.93    $  &  $29.78 \pm 4.88    $   &  $34.01 \pm 4.39    $   \\   
\hline                                                   
\end{tabular}
\end{table*}

\section{Analyses of anomalies} \label{sec:four}

A planetary lens corresponds to a binary lens with a very low mass ratio between the lens
components. In order to explain the observed anomalies in the lensing events, we conduct a 2L1S
modeling. Assuming a rectilinear relative motion between the lens and source, the behavior of a
2L1S event is characterized by seven basic parameters. The first three parameters $(t_0, u_0, \te)$
describe the source's approach to the lens: $t_0$ represents the time of the closest lens-source
approach, $u_0$ is the impact parameter of the approach, and $\te$ denotes the event time scale.
The 2L lens system is characterized by two additional parameters $(s,q)$, where $s$ denotes the 
projected separation (scaled to $\thetae$) and $q$ represents the mass ratio between the lens 
components. The parameter $\alpha$ indicates the incidence angle of the source relative to the 
binary-lens axis.  The last parameter, $\rho$, defined as the ratio of the angular source radius 
$\theta_*$ scaled to $\thetae$, characterizes the deformation of the lensing light curve during 
planetary perturbations by finite-source effects \citep{Bennett1996}.

Through modeling, we determined the lensing solution comprising a set of parameters that best
describe the observed anomalies. In this procedure, we searched for the binary parameters $(s, q)$
using a grid approach with multiple initial values of $\alpha$, while the remaining parameters 
were determined through a downhill approach using the Markov chain Monte Carlo (MCMC) algorithm.
Throughout this process, we utilized the map-making method \citep{Dong2006}, an improved version 
of the ray-shooting method, to compute finite magnifications. Additionally, we examined the
$\chi^2$ map on the grid parameter plane to assess the potential existence of degenerate
solutions.  In the concluding phase, we fine-tuned the parameters of the identified local 
solution, allowing for variability.

It is known that planetary signals arising from major-image perturbations can often be 
mimicked by a subset of binary-source (1L2S) events with extreme flux ratios between the 
source stars \citep{Gaudi1998, Gaudi2004}. To assess the degeneracy between 2L1S and 1L2S 
interpretations for each event, we conduct additional modeling under the 1L2S scenario.  
Modeling a 1L2S event requires incorporating three additional parameters $(t_{0,2}, u_{0,2}, 
q_F)$ alongside the parameters $(t_0, u_0, \te)$ of a 1L1S model, to address the anomalies 
resulting from the presence of a companion ($S_2$) to the primary source ($S_1$) \citep{Hwang2013}.  
Here, $(t_{0,2}, u_{0,2})$ designate the closest approach time and impact parameter of $S_2$, 
while $q_F$ represents the flux ratio between the companion and primary source stars.  In cases 
for which the lens passes over either of the source stars, additional parameters $\rho_1$ and/or 
$\rho_2$ should be included in the modeling.  In the 1L2S modeling process, we initially determine 
the parameters of the 1L1S solution by analyzing the data while excluding those in the region 
around the anomaly.  Subsequently, we ascertain the binary-source parameters, taking into account 
the location and magnitude of the anomaly.  In the following subsections, we present the analyses 
of the individual events.

\subsection{KMT-2021-BLG-2609} \label{sec:four-one}

The lensing event KMT-2021-BLG-2609 was found through the KMTNet survey during the later
stage of the bulge season on September 24, 2021, corresponding to the reduced Heliocentric
Julian date ${\rm HJD}^\prime\equiv {\rm HJD}-2450000 = 9481$.  The equatorial coordinates 
of the source are (RA, DEC)$_{\rm J2000}$ = (17:34:04.53, -27:35:48.98), which corresponds 
to the Galactic coordinates $(l,b)=(-0^\circ\hskip-2pt .2160, -2^\circ\hskip-2pt .8649)$.  
The source has a baseline magnitude $I_{\rm base}=18.48$ and the $I$-band extinction toward 
the field is $A_I = 3.29$.  The extinction is estimated as $A_I = 7\,A_K$, where the $K$-band 
extinction $A_K$ was adopted from \citet{Gonzalez2012}.  The source lies in the BLG15 field 
toward which observations were conducted with a 1-hour cadence.

\begin{figure}[t]
\includegraphics[width=\columnwidth]{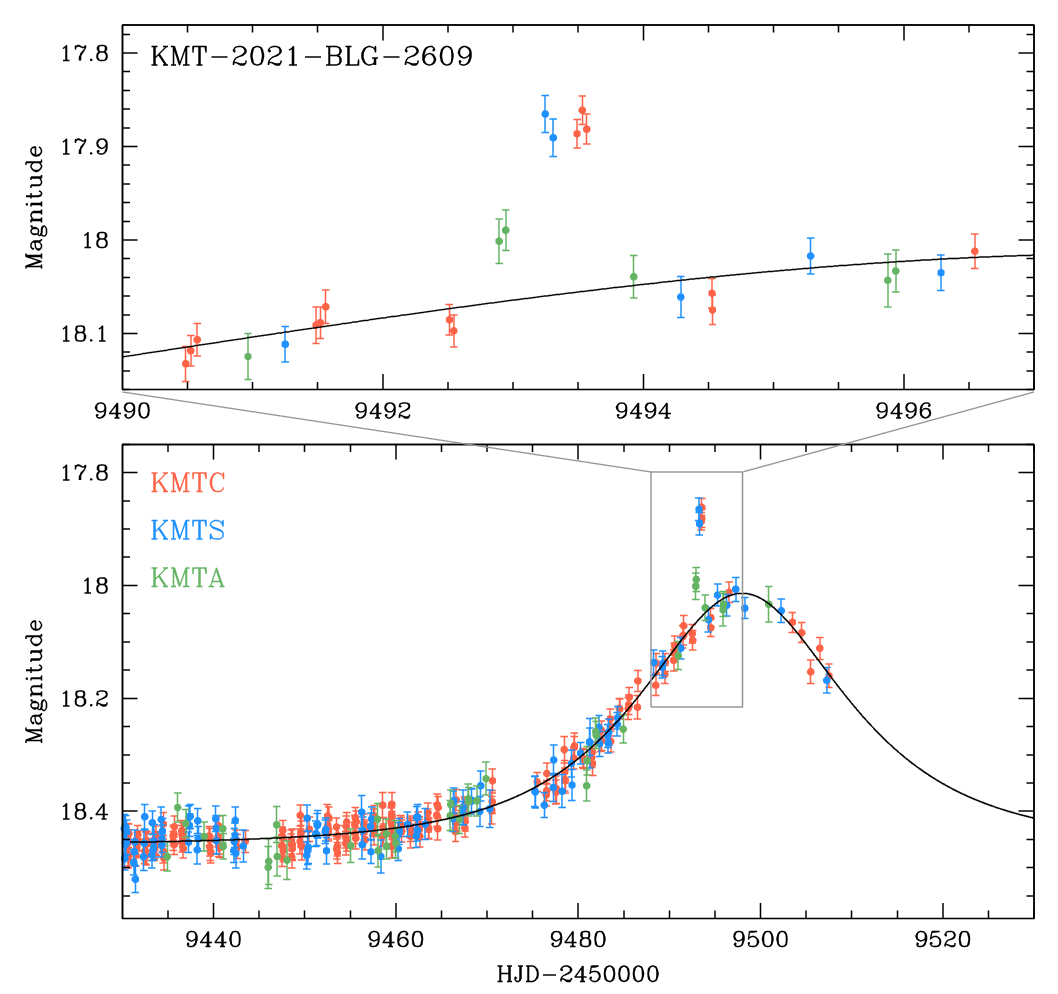}
\caption{
Light curve of the lensing event KMT-2021-BLG-2609. The lower panel shows the entire 
view, while the upper panel provides a close-up of the anomaly region. The colors of 
the data points match those in the legend, representing the telescopes employed for 
data acquisition. The solid curve drawn over the data points is a 1L1S model obtained 
by excluding the data around the anomaly.
}
\label{fig:two}
\end{figure}

\begin{figure}[t]
\includegraphics[width=\columnwidth]{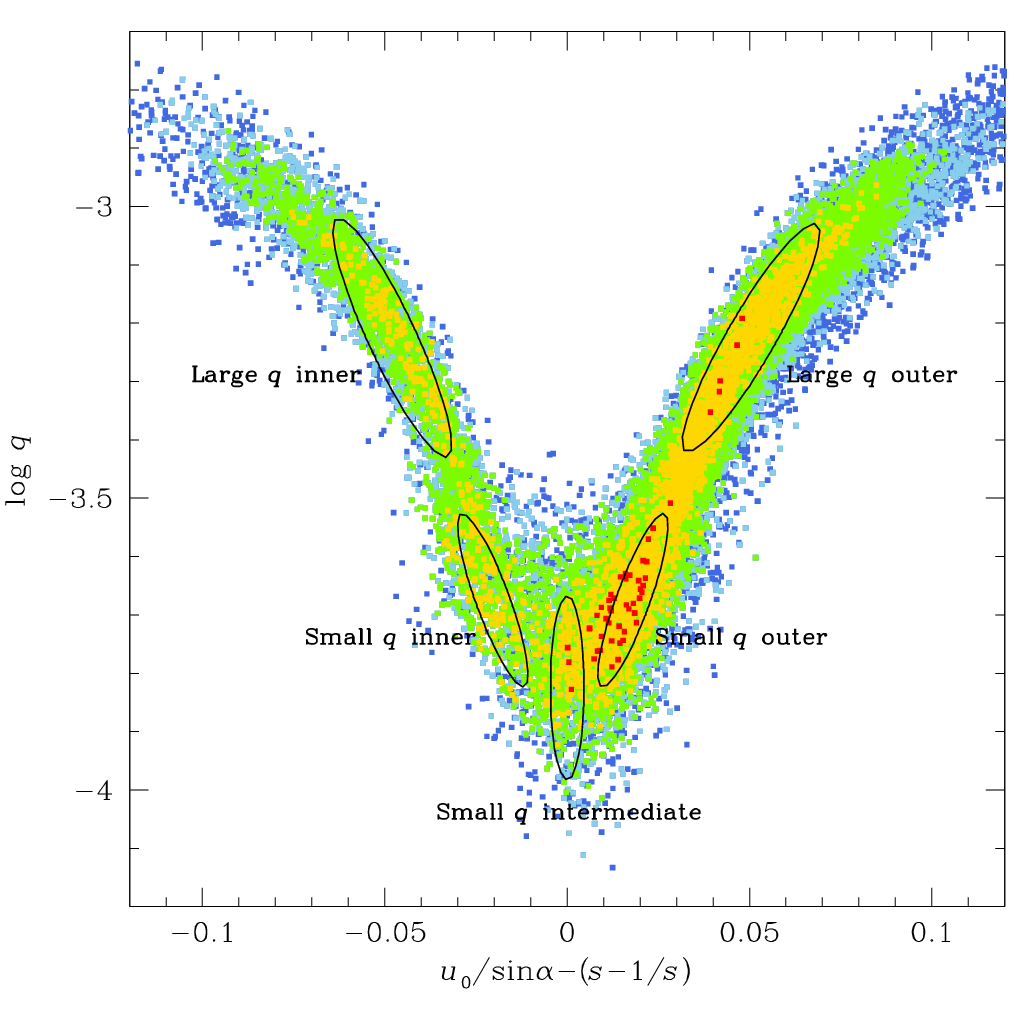}
\caption{
Scatter plot of points in the MCMC chain for the 2L1S model of KMT-2021-BLG-2609.  The 
ellipses mark the approximate locations of the five local solutions.  The color codeing 
is set to represent points with 
$\leq 1\sigma$ (red),
$\leq 2\sigma$ (yellow),
$\leq 3\sigma$ (green),
$\leq 4\sigma$ (cyan), and
$\leq 5\sigma$ (blue).
}
\label{fig:three}
\end{figure}

Figure~\ref{fig:two} displays the light curve of the lensing event KMT-2021-BLG-2609. 
Notably, there are no data points available after ${\rm HJD}^\prime \sim 9507$, marking 
the conclusion of the bulge season. The event reached its peak magnification on 
${\rm HJD}^\prime\sim 9497.9$, exhibiting a relatively low magnification of $A_{\rm peak} 
\sim 1.6$. Approximately four days before reaching its peak, the light curve exhibited a 
positive anomaly lasting for about two days. The upper panel of Figure~\ref{fig:two} provides 
a zoomed-in view of this anomaly region, which was captured by data obtained from all three 
KMTNet telescopes.

Two scenarios can explain a short-term positive anomaly in the light curve of a 
low-magnification event. The first scenario involves a planetary interpretation, in 
which the signal arises from the source nearing a planetary caustic positioned away 
from the location of the planet's host. The second scenario entails a 1L2S interpretation, 
in which the main light curve is generated by the primary source's approach to the lens 
with a large impact parameter, while the anomaly results from the close approach of a 
very faint companion source to the lens. We investigate both scenarios.

\begin{figure}[t]
\includegraphics[width=\columnwidth]{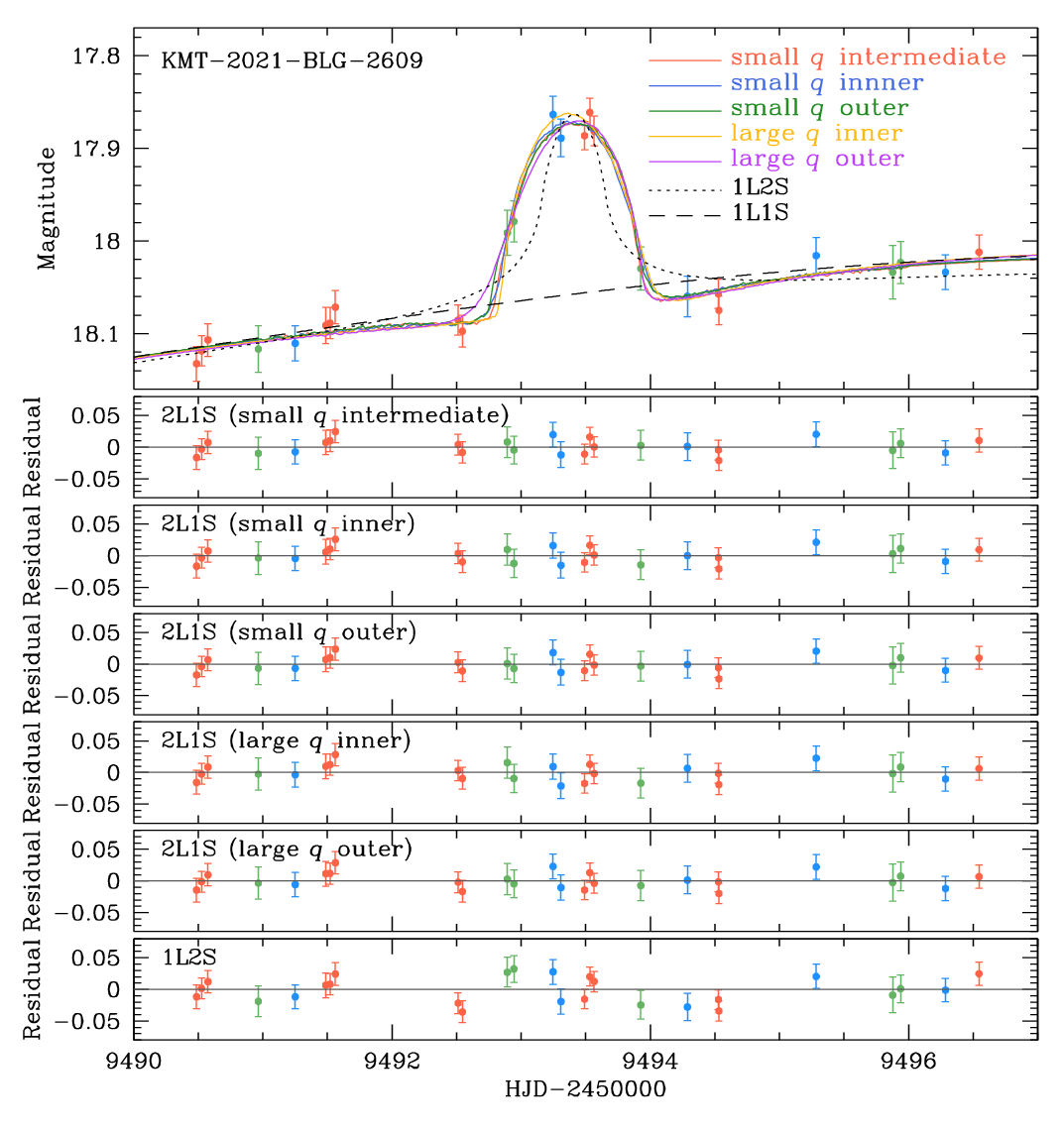}
\caption{
Zoomed-in view around the anomaly region of KMT-2021-BLG-2609. The data points are overlaid 
with the model curves representing the intermediate, inner, and outer 2L1S solutions, as well 
as the 1L1S and 1L2S solutions.  The residuals from the individual models are displayed in 
the lower panels.
}
\label{fig:four}
\end{figure}

\begin{figure}[t]
\includegraphics[width=\columnwidth]{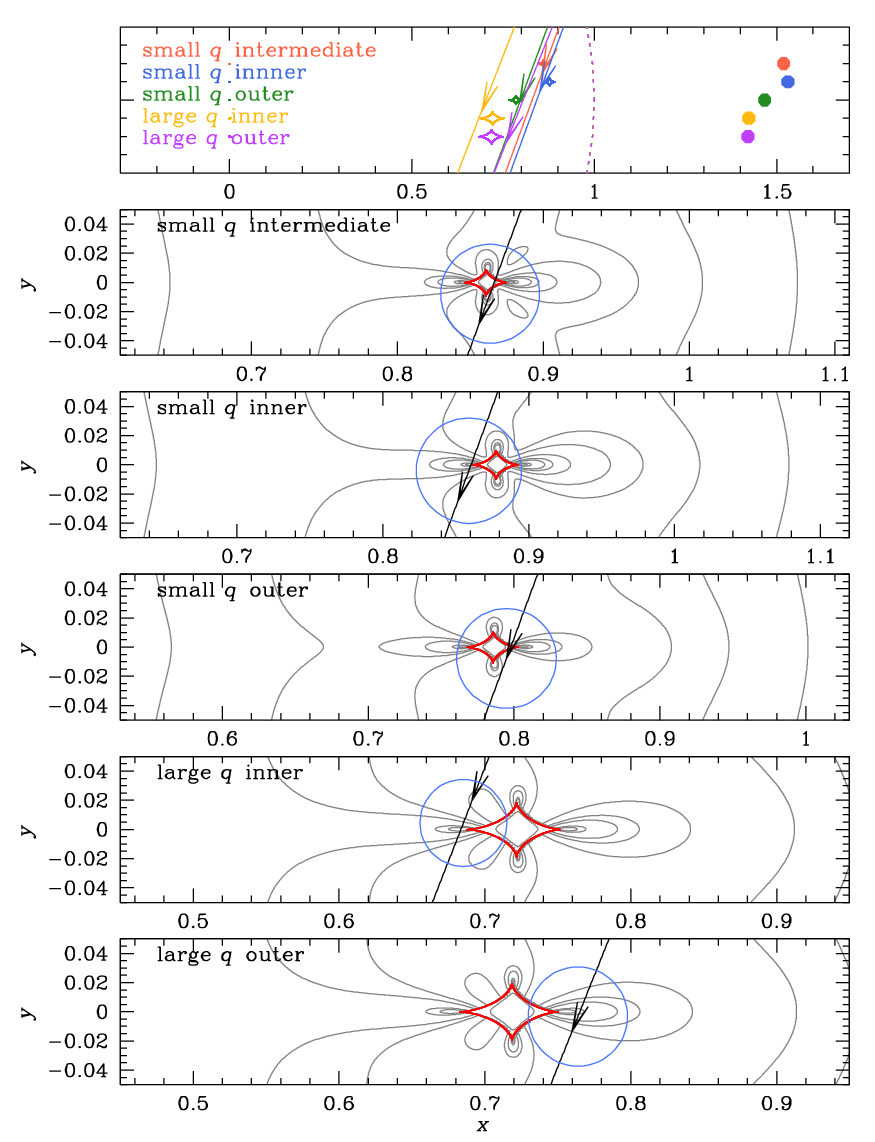}
\caption{
Lens-system configurations of the lensing event KMT-2021-BLG-2609 for the five local 
solutions.  The top panel provides a wide-angle perspective, showcasing the caustic 
and the position of the planet.  The planet is marked by a small filled dot.  Each of 
the lower panels presents a close-up view around the caustic. In these panels, the red 
figure indicates the caustic, while the arrowed line represents the source trajectory. 
The grey curves surrounding the caustic represent equi-magnification contours. Additionally, 
the blue circle positioned on the source trajectory represents the source whose size is 
adjusted to match that of the caustic.
}
\label{fig:five}
\end{figure}

From the modeling conducted under the planetary lens scenario, we have identified five 
distinct local solutions. In Figure~\ref{fig:three}, the positions of these individual 
solutions are depicted on the $\Delta\xi$--$\log q$ parameter plane, where $\Delta\xi= 
u_0/\sin\alpha -(s-1/s)$ represents the separation between the source and the center of 
the planetary caustic at the time of the planetary perturbation \citep{Hwang2018}.  Among 
these solutions, three exhibit smaller mass ratios falling within the range of $q \sim (0.15$ 
-- 0.22) $\times 10^{-3}$, while the remaining two solutions feature larger mass ratios 
$q\sim (0.53$ -- 0.58) $\times 10^{-3}$. We designate the former and latter solution groups 
as "small $q$" and "large $q$" solutions, respectively.

Further investigation reveals that the two solutions within the "large-$q$" group, with 
$\Delta\xi\sim -0.04$ and $+0.04$, originate from the inner--outer degeneracy initially 
identified by \citet{Gaudi1997}. These solutions are hence labeled as "large $q$ inner" 
and "large $q$ outer" solutions.  Similarly, the two solutions in the "small-$q$" group
with $\Delta\xi\sim -0.02$ and $+0.02$ also arise from this degeneracy, while the other 
solution has a separation $\Delta\xi\sim 0.0$.  We designate these solutions as "small $q$ 
inner", "small $q$ outer", and "small $q$ intermediate" solutions.  The degeneracy between 
the intermediate and the other solutions was extensively analyzed by \citet{Hwang2018}. They 
termed the solution as "Cannae" for a scenario in which the source entirely encompasses the 
planetary caustic, while they labeled the solution as "von Schlieffen" for a situation in 
which the source encompasses only one flank of the caustic.  According to this classification, 
the two large-$q$ solutions are von Schlieffen solutions, while all three of the remaining 
solutions, that is, small-$q$ solutions, are Cannae solutions. We note that, exactly as was 
the case for OGLE-2017-BLG-0173 \citep{Hwang2018}, the von Schlieffen solutions have 
substantially higher mass ratios than the Cannae solutions. We conjecture that this is a 
generic feature of the Cannae--von Schlieffen degeneracy, but we do not further pursue this 
question here.

In Table~\ref{table:one}, we present the lensing parameters for the solutions obtained 
under the 2L1S scenario, along with the corresponding $\chi^2$ values for the fits.
Figure~\ref{fig:four} illustrates the model curves for the five solutions and residuals 
from the models in the vicinity of the anomaly.  It is observed that the degeneracy among 
the solutions is highly pronounced, with $\chi^2$ 
differences among them being $\Delta\chi^2\leq 1.4$.

Figure~\ref{fig:five} illustrates the lens-system configurations of the individual local 
solutions. The top panel presents a broader view, including the planetary caustic and the 
position of the planet, while the lower panels depict a zoomed-in perspective around the 
caustic for the individual local solutions.  In all examined cases, the anomaly was interpreted 
through the source approach to the planetary caustic induced by a planet for which the projected 
planetary separation exceeded unity.  According to the inner solution, the source traversed the 
inner region of the planetary caustic relative to the planet's host, whereas according to the 
outer solution, it traversed the outer region. For the small-$q$ intermediate solution, the 
source center passed very close to the caustic center. In all instances, the source passed 
over the caustic, leading to the determination of the normalized source radius.  Notably, all 
solutions yield similar normalized source radii falling within the range of $\rho\sim (29.8$ 
-- $34.8)\times 10^{-3}$ in terms of a median value.

\begin{table}[t]
\caption{1L2S lensing parameters of KMT-2021-BLG-2609\label{table:two}}
\begin{tabular*}{\columnwidth}{@{\extracolsep{\fill}}lllcc}
\hline\hline
\multicolumn{1}{c}{Parameter}        &
\multicolumn{1}{c}{$\rho_2\neq 0$}   &
\multicolumn{1}{c}{$\rho_2=0$}       \\
\hline
 $\chi^2$                    &     $1726.7             $   &   $1741.0             $  \\   
 $t_0$ (HJD$^\prime$)        &     $9498.498 \pm 0.260 $   &   $9498.468 \pm 0.262 $  \\   
 $u_0$                       &     $1.088 \pm 0.239    $   &   $0.998 \pm 0.276    $  \\   
 $\te$ (days)                &     $13.74 \pm 1.93     $   &   $14.53 \pm 2.33     $  \\   
 $t_{0,2}$ (HJD$^\prime$)    &     $9493.401 \pm 0.032 $   &   $9493.414 \pm 0.021 $  \\   
 $u_{0,2}$ (10$^{-3}$)       &     $-0.23 \pm 6.25     $   &   $9.47  \pm 2.80     $  \\   
 $\rho_2$ (10$^{-3}$)        &     $19.73 \pm 4.62     $   &   --                     \\   
 $q_F$ (10$^{-3}$)           &     $1.78  \pm 0.74     $   &   $2.62  \pm 1.11     $  \\   
\hline                                                   
\end{tabular*}
\end{table}

It is known that the planetary separations of the pair of solutions subject to the inner--outer 
degeneracy follow the relationship established by \citet{Hwang2022} and \citet{Gould2022b}:
\begin{equation}
s_\pm^\dagger = \sqrt{s_{\rm inner} \times s_{\rm outer}   } = 
{\sqrt{u^2_{\rm anom} + 4} \pm u_{\rm anom}\over 2}.
\label{eq2}
\end{equation}
Here $u_{\rm anom}$ represents the lens-source separation at the time of the anomaly 
($t_{\rm anom}$), given by:
\begin{equation}
u_{\rm anom} = \sqrt{u_0^2+\tau_{\rm anom}^2};\qquad
\tau_{\rm anom} = { t_{\rm anom}-t_0 \over \te}.
\label{eq3}
\end{equation}
The symbols "$+$" and "$-$" in the first and last terms of Eq.~(\ref{eq2}) indicate the 
perturbations corresponding to the major and minor images, respectively.  For the pair of 
inner and outer solutions in the small $q$ group with $(t_0, u_0, \te, t_{\rm anom}) \sim 
(9498.0, 0.78, 16.0, 9493.5)$
\footnote{ 
Here we take the mean values of the inner and outer solutions:
$\langle t_0 \rangle = (t_{0,{\rm inner}}+t_{0,{\rm outer}})/2$,
$\langle u_0 \rangle = (u_{0,{\rm inner}}+u_{0,{\rm outer}})/2$, 
$\langle \te \rangle = (t_{\rm E,inner}+t_{\rm E,outer})/2$.
}, 
we find that $s_+^\dagger =1.497$, closely aligning with the geometric mean of the planetary 
separations, $(s_{\rm inner} \times s _{\rm outer})^{1/2} \sim 1.492$.  Similarly, for the pair 
of inner and outer solutions in the large $q$ group with $(t_0, u_0, \te, t_{\rm anom}) \sim 
(9497.9, 0.67, 17.4, 9493.5)$, we determine that $s_+^\dagger =1.420$, which again closely aligns 
with the geometric mean of the planetary separations, $(s_{\rm inner} \times s_{\rm outer})^{1/2} 
\sim 1.423$.  The fact that the lensing parameters of the two solutions conform to the analytic 
relation derived from investigating the lensing properties of the inner and outer solutions 
suggests that the pair of solutions arises from the inner--outer degeneracy.

\begin{figure}[t]
\includegraphics[width=\columnwidth]{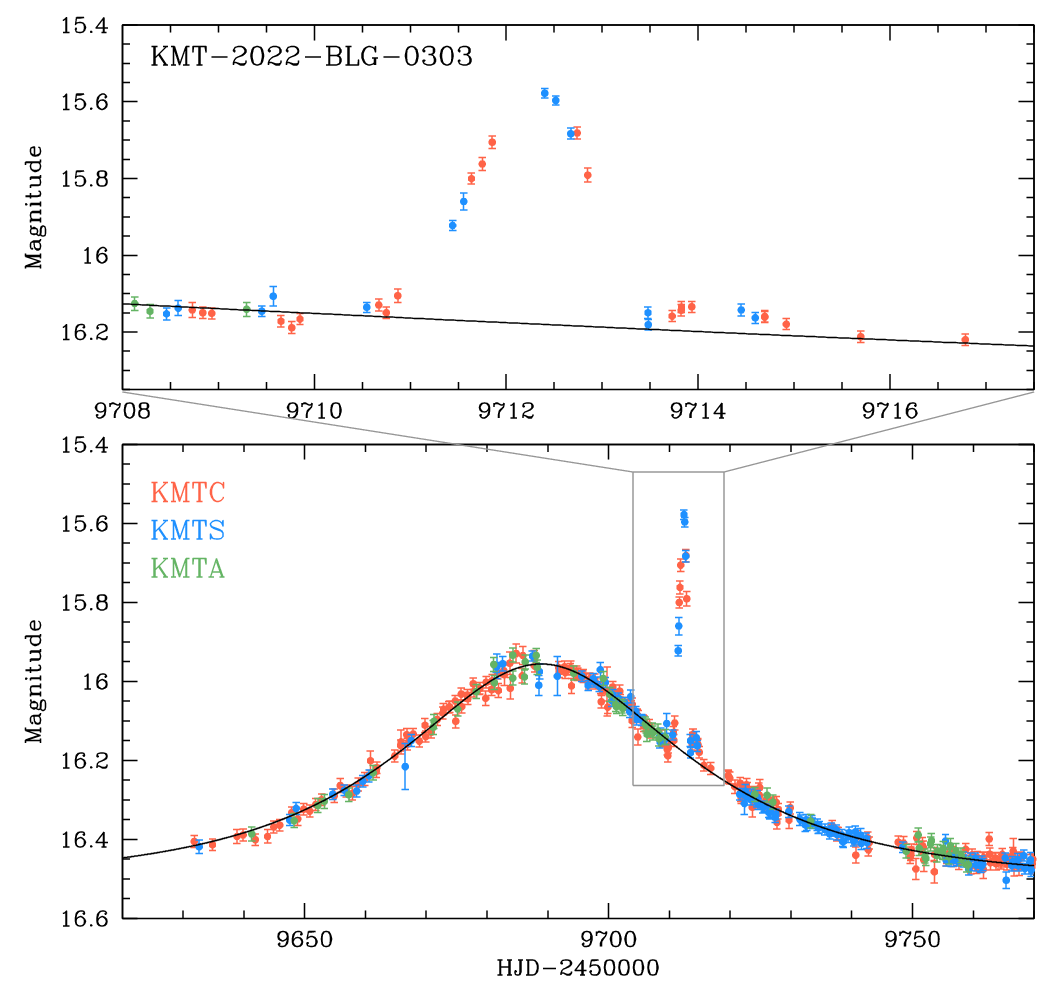}
\caption{
Lensing light curve of KMT-2022-BLG-0303.  The notations used are same as those presented in
Fig.~\ref{fig:two}.
}
\label{fig:six}
\end{figure}

We find that the 1L2S model is less favored for two major reasons, even though  it approximately 
delineates the anomaly.  Firstly, the 1L2S model provides a worse fit compared to the planetary 
solution, with a $\Delta\chi^2\sim 23$.  The lensing parameters for the 1L2S solution are provided 
in Table~\ref{table:two}, and the model curve and residual are depicted in Figure~\ref{fig:four}.  
Notably, the 1L2S model fails to accurately describe the slight negative deviations just before 
and after the main bump of the anomaly, which are well-captured by the planetary model.  Secondly, 
the 1L2S parameters are physically implausible.  In Section~\ref{sec:five}, we will show that the 
de-reddened magnitude of the primary source for the 1L2S solution is $I_{S_1,0}=14.40$, that is, 
similar to the 2L1S solution.  Therefore, the magnitude of the secondary source is $I_{S_2,0} = 
I_{S_1,0} - 2.5 \log(q_F) = 21.3$, which corresponds to an early M dwarf with a physical radius 
$R_2 \simeq 0.5\,M_\odot$, which corresponds to an angular radius $\theta_{*,2}\simeq 0.3\,\mu$as.  
These numbers imply $\theta_\e \simeq 15\,\mu$as, and so a relative lens-source proper motion 
$\mu\simeq 0.4$~mas/yr.  According to Equations (22) and (23) of \citet{Gould2022a}, the probability 
of such a low proper motion (for events consistent with a planetary interpretation) is 
$p=(\mu/6~{\rm mas~yr}^{-1}) \sim 0.0045$.  Combined with the fact that this solution is also 
disfavored by $\Delta\chi^2=23.4$, this low $p$-value implies that this solution is effectively 
ruled out.  However, because $\rho_2$ is relatively weakly constrained, due diligence requires 
that we consider the possibility that $\rho_2$ is much smaller than the best fit value given in 
Table~\ref{table:two}, which would eliminate the low-proper-motion argument. Therefore, in the 
final column of Table~\ref{table:two}, we report results for a 1L2S solution with $\rho_2$ fixed 
at $\rho_2=0$. Table~\ref{table:two} shows that such a solution has yet higher $\chi^2$, that is, 
$\Delta\chi^2=37.6$, showing that such solutions are also ruled out.  We therefore find that this 
event is securely 2L1S.  Given that the 2L1S model adequately explains the observed anomaly, we do 
not consider more complex models such as the binary-lens binary-source (2L2S) model, as illustrated 
by the planetary event KMT-2021-BLG-1898 \citep{Han2022}, or the triple-lens (3L1S) model, as 
illustrated by the planetary event OGLE-2023-BLG-0836 \citep{Han2024d}.

\subsection{KMT-2022-BLG-0303} \label{sec:four-two}

The microlensing event KMT-2022-BLG-0303 was first identified on April 4, 2022, corresponding 
to ${\rm HJD}^\prime \sim 9673$.  The equatorial and Galactic coordinates of the source are 
(RA, DEC)$_{\rm J2000}$ = (17:52:07.95, -22:41:20.69) and $(l,b)= (+6^\circ\hskip-2pt .1105, 
-1^\circ\hskip-2pt .9615)$, respectively.  The $I$-band baseline magnitude of the source is 
$I_{\rm base}=16.48$ and the extinction toward the field is $A_I = 1.64$.  The source lies in 
the BLG38 field and observations were done with a 2.5-hour cadence.

The event reached its maximum magnification at around 
${\rm HJD}^\prime \sim 9689.2$, with a relatively low magnification of $A_{\text{peak}} \sim 1.7$.  
Figure~\ref{fig:six} illustrates the light curve of the event, revealing a prominent anomaly 
appearing in the falling side of the light curve centered around ${\rm HJD}^\prime \sim 9713$. 
The anomaly was covered through the integration of the KMTC and KMTS datasets, revealing a 
distinctive dual bump pattern: a prominent peak centered around ${\rm HJD}^\prime \sim 9712.3$ 
and a secondary peak around ${\rm HJD}^\prime \sim 9714.2$

\begin{figure}[t]
\includegraphics[width=\columnwidth]{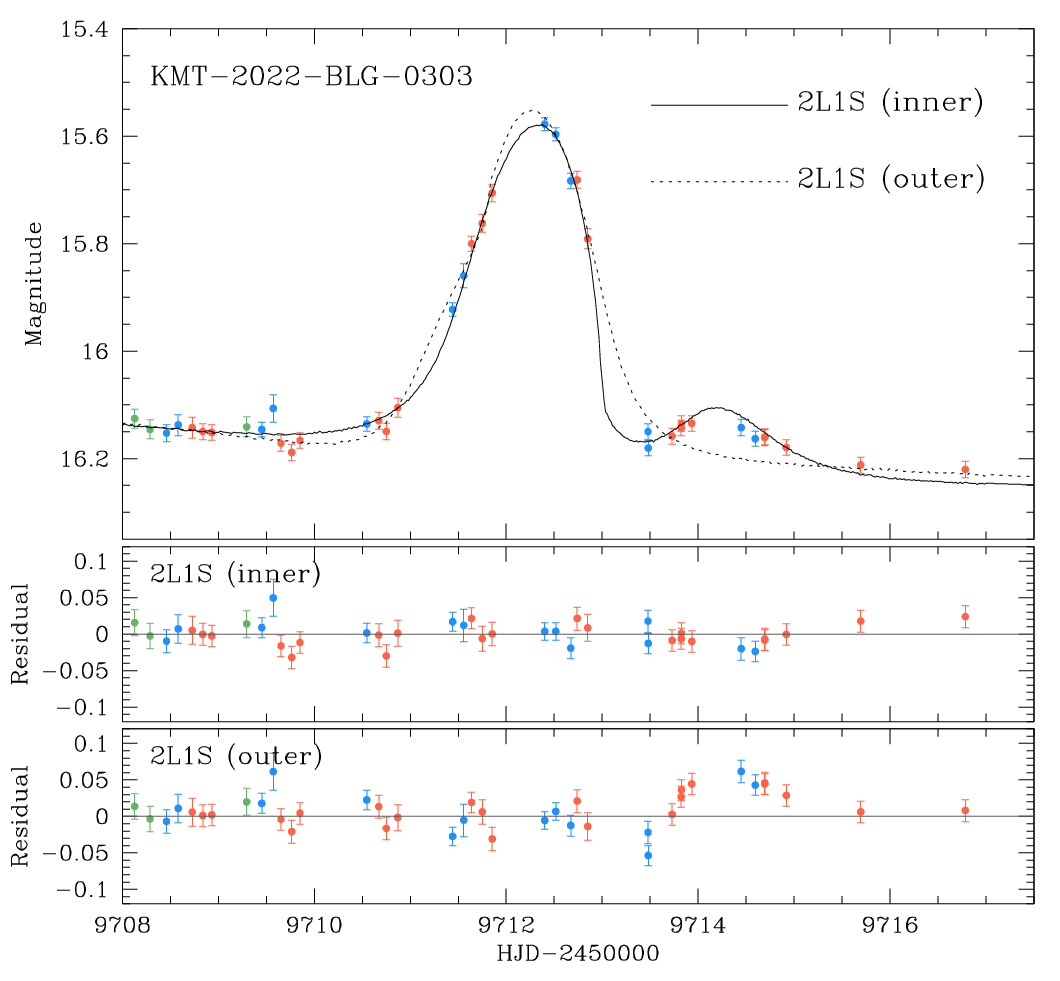}
\caption{
Zoomed-in view around the anomaly region of KMT-2022-BLG-0303. The model curves of the inner 
and outer 2L1S solutions and their residuals are presented.
}
\label{fig:seven}
\end{figure}

From the 2L1S modeling, we identified a pair of local planetary solutions with $(s, q)_{\rm inner}
\sim (1.61, 1.33\times 10^{-3})$ and $(s, q)_{\rm outer}\sim(1.60, 0.58\times 10^{-3})$. These 
solutions arise from the inner--outer degeneracy. The complete lensing parameters for both 
solutions are presented in Table~\ref{table:three}, and the model curves around the region of 
the anomaly are depicted in Figure~\ref{fig:seven}.  Unlike the previous event KMT-2021-BLG-2609, 
the degeneracy in KMT-2022-BLG-0303 is significantly less severe, and the inner solution is 
strongly favored over the outer solution, with a substantial $\chi^2$ difference of $\Delta\chi^2 
= 79.2$.

\begin{figure}[t]
\includegraphics[width=\columnwidth]{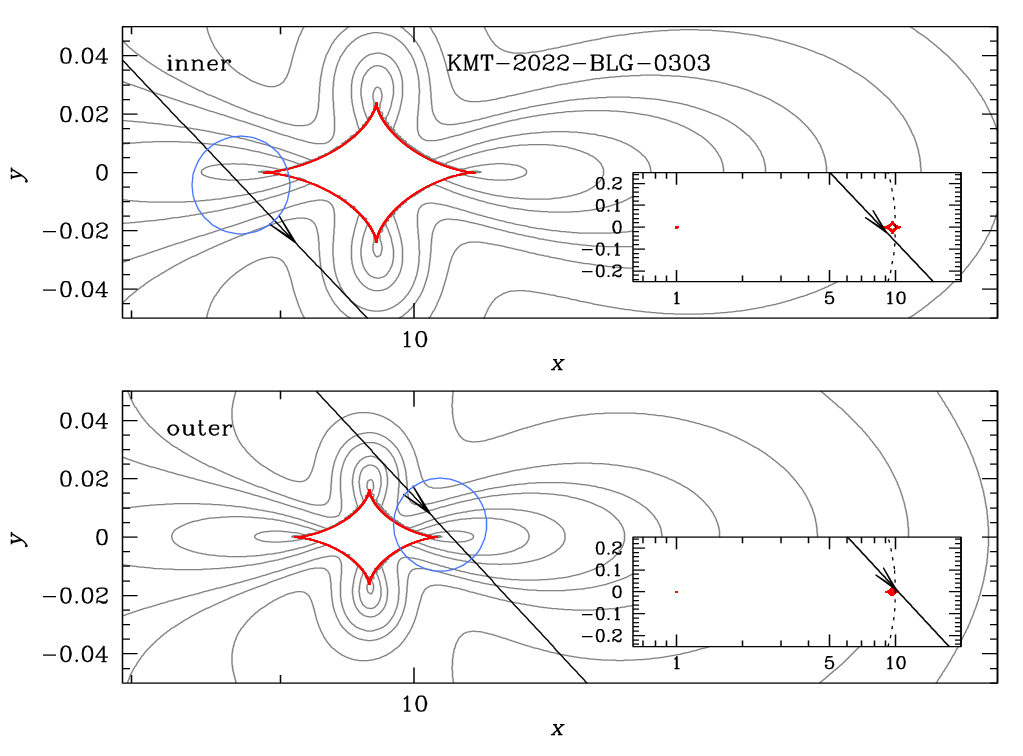}
\caption{
Configurations of KMT-2022-BLG-0303 lens system. The notations correspond to those utilized in 
Fig.~\ref{fig:five}.
}
\label{fig:eight}
\end{figure}

\begin{table}[t]
\caption{Lensing parameters of KMT-2022-BLG-0303 \label{table:three}}
\begin{tabular*}{\columnwidth}{@{\extracolsep{\fill}}lllcc}
\hline\hline
\multicolumn{1}{c}{Parameter}    &
\multicolumn{1}{c}{Inner}        &
\multicolumn{1}{c}{Outer}        \\
\hline
  $\chi^2$               &  $566.0             $ &  $645.2             $   \\   
  $t_0$ (HJD$^\prime$)   &  $9689.154 \pm 0.112$ &  $9689.178 \pm 0.122$   \\   
  $u_0$                  &  $0.680 \pm 0.019   $ &  $0.735 \pm 0.015   $   \\   
  $\te$ (days)           &  $35.71 \pm 0.66    $ &  $34.02 \pm 0.53    $   \\   
  $s$                    &  $1.609 \pm 0.016   $ &  $1.602 \pm 0.013   $   \\   
  $q$ (10$^{-3}$)        &  $1.330 \pm 0.053   $ &  $0.581 \pm 0.022   $   \\   
  $\alpha$ (rad)         &  $3.9536 \pm 0.0065 $ &  $3.9618 \pm 0.0060 $   \\   
  $\rho$ (10$^{-3}$)     &  $16.75 \pm 0.51    $ &  $15.79 \pm 0.40    $   \\   
\hline                                                   
\end{tabular*}
\end{table}

\begin{table*}[t]
\caption{Source parameters and spectral types\label{table:four}}
\begin{tabular}{llllll}
\hline\hline
\multicolumn{1}{c}{Quantity}            &
\multicolumn{1}{c}{KMT-2021-BLG-2609}   &
\multicolumn{1}{c}{KMT-2022-BLG-0303}   \\
\hline
$(V-I, I_{\rm S})$            &  $(3.976 \pm 0.115, 17.846 \pm 0.012)$  & $(2.906 \pm 0.071, 16.753 \pm 0.043) $  \\
$(V-I, I)_{\rm RGC}$          &  $(3.926, 17.699) $  & $(2.939, 15.984) $  \\
$(V-I, I)_{{\rm RGC},0}$      &  $(1.060, 14.456) $  & $(1.060, 14.273) $  \\
$(V-I, I)_{{\rm S},0}$        &  $(1.110 \pm 0.115, 14.603 \pm 0.012) $  & $(1.027 \pm 0.071, 15.042 \pm 0.043) $  \\
Source type                   &   K3III              &  K3III              \\
\hline
\end{tabular}
\end{table*}

The reason for resolving the degeneracy became apparent upon scrutinizing the configuration 
of the lens system illustrated in Figure~\ref{fig:eight}.  Similar to the case of KMT-2021-BLG-2609, 
the configuration suggests that the anomaly originated from the close approach of the source to the 
planetary caustic, induced by the presence of the wide planet.  While the caustic configurations 
resemble those of KMT-2021-BLG-2609, there is an important difference: the ratio of the source size 
to the caustic size for KMT-2022-BLG-0303 is substantially smaller than that for KMT-2021-BLG-2609.  
As a result, the anomaly reflects the structure of the caustic, leading to the observed dual-bump 
feature: the strong bump occurs when the source approaches the strong on-axis cusp of the caustic, 
while the weak bump arises when the source nears the weak off-axis cusp.  In contrast, the outer 
solution fails to generate this dual-bump feature because the source initially approaches the 
off-axis cusp.  Because the source encompasses one flank of the caustic, the normalized source 
radius was constrained to be $\rho = (16.75 \pm 0.51) \times 10^{-3}$.  The dual nature of the 
anomaly also facilitates the resolution of the degeneracy between the planetary and 1L2S 
interpretations, and it was found that the inner 2L1S model is favored over the 1L2S model by 
$\Delta\chi^2=435.4$.

\section{Source stars and angular Einstein radii} \label{sec:five}

In this section, we provide detailed specifications regarding the source stars involved in the 
events.  Identifying source stars is crucial not only for fully characterizing the events but 
also for determining the angular Einstein radius. The Einstein radius is computed using the relation
\begin{equation}
\thetae = {\theta_* \over \rho},
\label{eq4}
\end{equation}
where the angular source radius $\theta_*$ can be deduced from the source type and the normalized
source radius is measured from modeling.

\begin{figure}[t]
\includegraphics[width=\columnwidth]{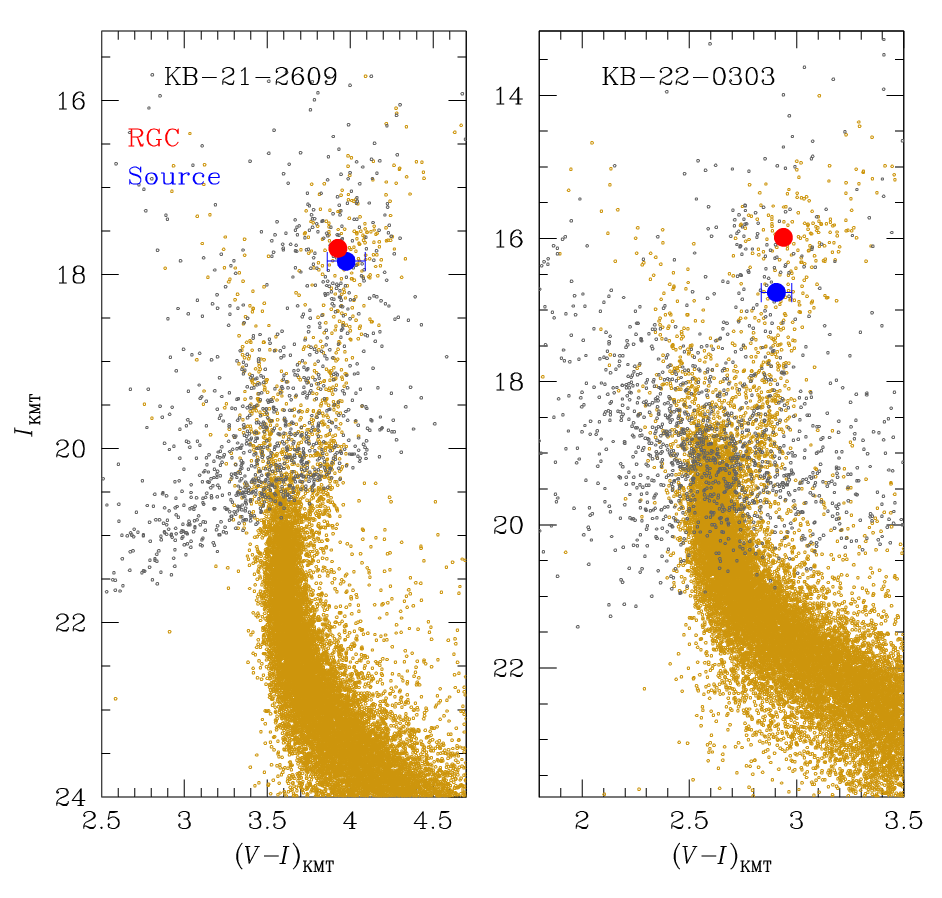}
\caption{
Source locations in the instrumental color-magnitude diagrams with respect to the centroids of 
red giant clump (RGC).  In each panel, grey and brown dots represent CMDs derived from KMTC and 
HST observations, respectively.
}
\label{fig:nine}
\end{figure}

We defined the type of the source by measuring its reddening and extinction-corrected (de-reddened)
color $(V-I)_{{\rm S},0}$ and brightness $I_{{\rm S},0}$. In the initial phase of this procedure, 
we estimated the instrumental color and magnitude $(V-I, I)_{\rm S}$ by conducting regression analysis 
on the data sets obtained from the pyDIA photometry code for both the $V$-band and $I$-band, relative 
to the model.  Figure~\ref{fig:nine} shows the locations of the source stars for the individual events 
on the color-magnitude diagrams (CMDs) of stars lying near the source stars.  For both events, obtaining 
reliable $V$-band source magnitude posed challenges because of either poor data quality or sparse coverage 
during the lensing magnification phase.  In this case, we determined the source color by averaging the 
colors of stars positioned on either the giant or main-sequence branches of the combined CMD, with 
$I$-band magnitude offsets from the centroid red giant clump (RGC) falling within the measured value's 
range.  The CMD was compiled by merging CMDs obtained from KMTC and {\it Hubble} Space Telescope (HST) 
observations.  The HST CMD data were derived from star observations in Baade's Window conducted by 
\citet{Holtzman1998}, and alignment of the KMTC and HST CMDs was achieved by utilizing the centroids 
of the RGC in the respective CMDs.

\begin{table}[t]
\caption{Einstein radii and relative lens-source proper motions\label{table:five}}
\begin{tabular*}{\columnwidth}{@{\extracolsep{\fill}}lllcc}
\hline\hline
\multicolumn{1}{c}{Quantity}            &
\multicolumn{1}{c}{KMT-2021-BLG-2609}   &
\multicolumn{1}{c}{KMT-2022-BLG-0303}   \\
\hline
$\theta_*$ ($\mu$as)          &  $5.992 \pm 0.804 $  & $4.517 \pm 0.450 $  \\
$\thetae$ (mas)               &  $0.177 \pm 0.032$   & $0.270 \pm 0.028 $  \\
$\mu$ (mas/yr)                &  $4.136 \pm 0.804$   & $2.758 \pm 0.292 $  \\
\hline
\end{tabular*}
\end{table}

In the second phase, we calibrated the source color and magnitude. This calibration was performed
using the approach introduced by \citet{Yoo2004}. In this methodology, the RGC centroid served
as a calibration reference because of its well-established de-reddened color and magnitude. By
measuring the offsets of the source in color and magnitude from the RGC centroid, denoted as
$\Delta(V-I, I)_{\rm S}$, we calculated the de-reddened color and magnitude of the source as
\begin{equation}
(V-I, I)_{{\rm S},0} = (V-I, I)_{{\rm RGC},0} + \Delta(V-I, I)_{\rm S}. 
\label{eq5}
\end{equation}
Here, $(V-I, I)_{{\rm RGC},0}$ represents the de-reddened color and magnitude of the RGC centroid. 
We adopted the $(V-I)_{{\rm RGC},0}$ value from \citet{Bensby2013} and the $I_{{\rm RGC},0}$ value 
from \citet{Nataf2013}. In Table~\ref{table:four}, we list the values of $(V-I, I)_{\rm S}$, 
$(V-I, I)_{\rm RGC}$, $(V-I, I)_{{\rm RGC},0}$, and $(V-I, I)_{{\rm S},0}$, for the analyzed events. 
Upon analyzing the determined colors and magnitudes, it was observed that the source stars in both 
events are K-type giants.

In the third phase, we derived the angular source radius from the measured source color and 
magnitude, and then estimated angular Einstein radius. To derive $\theta_*$, we initially 
transformed the $V-I$ color into $V-K$ color using the \citet{Bessell1988} relation. Next, we 
determined $\theta_*$ based on the \citet{Kervella2004} relationship between the $(V-K, I)$ 
and the angular stellar radius. With the estimated source radius, the angular Einstein radius 
was measured using the relation in Eq.~(\ref{eq4}), and the relative lens-source proper motion 
was calculated as 
\begin{equation} \mu = { \thetae\over \te}.
\label{eq6}
\end{equation}
In Table~\ref{table:five}, we list the estimated values of the angular source radius, Einstein 
radius, and relative proper motion  for the individual events.  For KMT-2021-BLG-2609, we computed 
estimates for $\thetae$ and $\mu$ based on the $\rho$ and $\te$ values obtained from the small-$q$ 
intermediate solution, which provides the best fit to the data. Notably, the alternative solutions 
yield $\te$ and $\rho$ values that are comparable, resulting in $\thetae$ and $\mu$ estimates 
similar to the presented values.

\begin{table}[t]
\caption{Physical lens parameters\label{table:six}}
\begin{tabular*}{\columnwidth}{@{\extracolsep{\fill}}lllcc}
\hline\hline
\multicolumn{1}{c}{Quantity}            &
\multicolumn{1}{c}{KMT-2021-BLG-2609}  &
\multicolumn{1}{c}{KMT-2022-BLG-0303}  \\
\hline
$M_{\rm host}$ ($M_\odot$)       & $0.20^{+0.27}_{-0.11}$                  &  $0.368^{+0.294}_{-0.197} $   \\ [0.4ex]
$M_{\rm planet}$ ($M_{\rm J}$)   & $0.032^{+0.044}_{-0.018}$\ (small $q$)  &  $0.513^{+0.410}_{-0.275} $   \\ [0.4ex]
\hskip50pt                       & $0.112^{+0.152}_{-0.061}$\ (large $q$)  &  --                           \\ [0.4ex]
$\dl$ (kpc)                      & $7.48^{+1.05}_{-1.17}$                  &  $6.57^{+0.89}_{-1.03} $      \\ [0.4ex]
$\ds$ (kpc)                      & $9.09^{+1.07}_{-1.04}$                  &  $8.10^{+0.91}_{-0.88} $      \\ [0.4ex]
$a_\perp$ (AU)                   & $2.30^{+0.32}_{-0.36}$                  &  $4.81^{+0.65}_{-0.76} $      \\ [0.4ex]
$p_{\rm disk}$                   &  21\%                                   &  28\%                         \\ [0.4ex]
$p_{\rm bulge}$                  &  79\%                                   &  72\%                         \\ [0.4ex]
\hline
\end{tabular*}
\end{table}

\section{Physical parameters of planetary systems} \label{sec:six}

In this section, we estimate the masses of the planet and host, along with the distance for 
each individual planetary system.  The physical parameters were determined based on the lensing 
observables, including the event time scale and the angular Einstein radius.  These observables 
are related to the mass $M$ and distance $\dl$ by the relations
\begin{equation}
\te = {\thetae \over \mu};\qquad 
\thetae = (\kappa M \pi_{\rm rel})^{1/2},
\label{eq7}
\end{equation}
where $\kappa = 4G/(c^2{\rm AU}) \simeq 8.14~{\rm mas}/M_\odot$, $\pi_{\rm rel} = \pi_{\rm L} 
- \pi_{\rm S} = {\rm AU}(1/\dl - 1/\ds)$ represents the relative lens-source parallax, and 
$\ds$ denotes the distance to the source. Besides these observables, the physical lens parameters 
can be more precisely determined by measuring an additional observable of the micro-lens parallax 
$\pie$, with which the mass and distance can be uniquely determined using the \citet{Gould2000} 
relations:
\begin{equation}
M={ \thetae \over \kappa \pie};\qquad
\dl = {{\rm AU} \over \pie\thetae + \pi_{\rm S} }.
\label{eq8}
\end{equation}
For none of the events could the microlens parallax be measured, either because of the relatively 
short time scales or because of the incomplete light curve coverage.

To estimate the mass and the distance to the lens, we conducted Bayesian analyses. 
These analyses integrated priors derived from the physical and dynamical distribution, as 
well as the mass function of lens objects, along with the constraints provided by $\te$ and $\thetae$.
In the analysis, we initially generated a large number of artificial lensing events using a Monte 
Carlo simulation. In the simulations, we allocated the locations of the lens and source, as well 
as their relative proper motion, based on a Galactic model. The mass of the lens was assigned using 
a mass function model. Specifically, we utilized the Galactic model proposed by \citet{Jung2021} 
and the mass-function model introduced by \citet{Jung2018}. Using the provided values of $(M, \dl, 
\ds, \mu)$, we then calculated the lensing observables $(t_{{\rm E},i}, \theta_{{\rm E},i})$ for 
each artificial event by applying the relations given in Eq.~(\ref{eq7}). Finally, we obtained 
posteriors of $M$ and $\dl$ by constructing their weighted distributions of simulated events. The 
weight assigned to each event was computed as
\begin{equation}
w_i = \exp\left(-{\chi_i^2 \over 2} \right);\qquad
\chi_i^2 = 
\left[{t_{{\rm E},i}-\te \over \sigma(\te)} \right]^2 + 
\left[{\theta_{{\rm E},i}-\thetae \over \sigma^2(\thetae}\right]^2,
\label{eq9}
\end{equation}     
where $[\te \pm \sigma(\te), \thetae \pm \sigma(\thetae)]$ denote the measured values and 
uncertainties for the lensing observables.

\begin{figure}[t]
\includegraphics[width=\columnwidth]{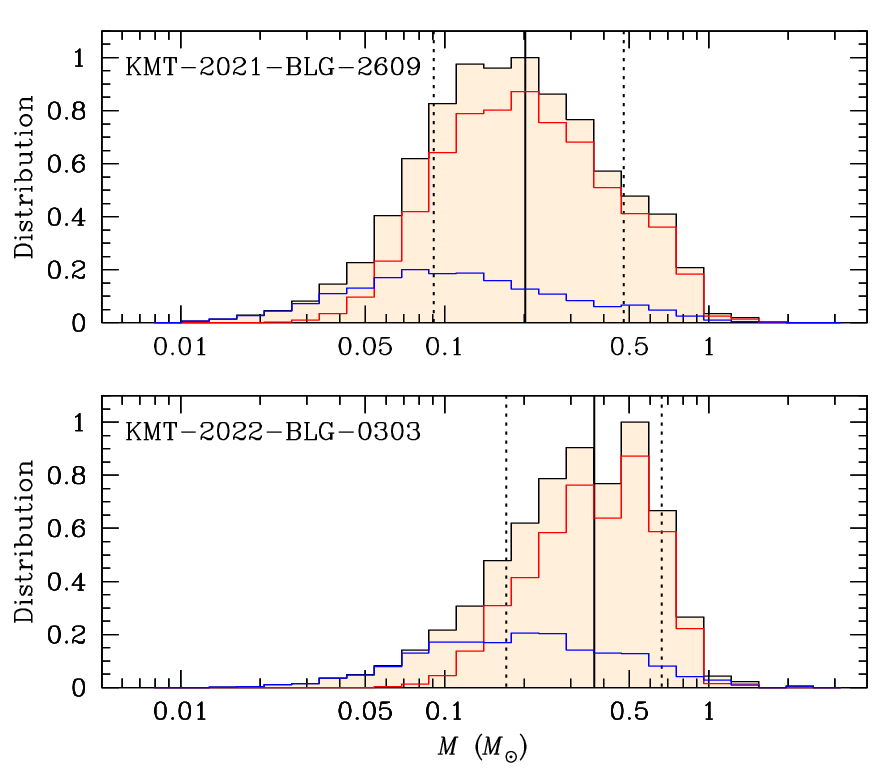}
\caption{
Bayesian posteriors for the masses of the planetary systems' host stars. Within each panel, 
the blue and red curves represent the contributions from the disk and bulge lens populations, 
respectively. The combined contribution of both populations is depicted by the black curve. 
The solid vertical line indicates the median value for each distribution, while the two dotted 
vertical lines denote the 1$\sigma$ uncertainty range.  
}
\label{fig:ten}
\end{figure}

\begin{figure}[t]
\includegraphics[width=\columnwidth]{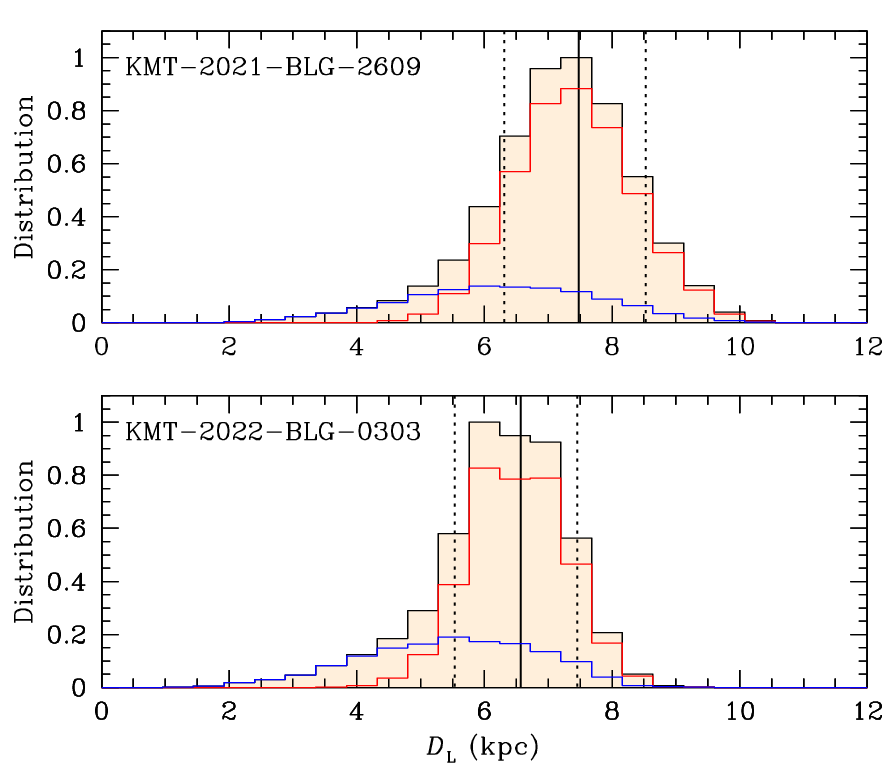}
\caption{
Bayesian posteriors for the distances to the planetary systems. The notation system employed 
here is consistent with that of Fig.~\ref{fig:ten}.	
}
\label{fig:eleven}
\end{figure}

Figures~\ref{fig:ten} and \ref{fig:eleven} illustrate the posterior distributions of mass and 
distance for the planetary systems. Within each distribution, we indicate the median value using 
a vertical solid line, while the uncertainty range is delineated by two dotted lines.  
Table~\ref{table:six} provides a summary of parameters including host mass ($M_{\rm host}$), 
planet mass ($M_{\rm planet}$), distances to the lens ($\dl$) and source ($\ds$), and projected 
planet-host separation ($a_\perp$).  The source distances were also derived from the Bayesian 
analyses based on the Galatic model.  For each parameter, the upper and lower limits are defined 
as the 16th and 84th percentiles of the Bayesian posterior.  The projected separation was estimated 
from the relation $a_\perp = s\thetae\dl$.  For KMT-2021-BLG-2609L, we present two values of the 
planet mass corresponding to the small and large $q$ solutions.

For the planetary system KMT-2021-BLG-2609L, the host is identified as a low-mass star with a mass 
of $\sim 0.20~M_\odot$. The planet's mass varies depending on the solution: it is $\sim 0.032~M_{\rm J}$ 
according to the small $q$ solution and $\sim 0.112~M_{\rm J}$ according to the large $q$ solution.
For the system KMT-2022-BLG-0303L, it features a planet with a mass $\sim 0.51~M_{\rm J}$ and a host 
star with a mass $\sim 0.37~M_\odot$.  The estimated distances are $\sim 7.5$~kpc for KMT-2021-BLG-2609L 
and $\sim 6.6$~kpc for KMT-2022-BLG-0303L.  Table~\ref{table:six}  presents the probabilities for the 
lens to be located in the disk ($p_{\rm disk}$) and bulge ($p_{\rm bulge}$). For both events, the 
probability of the lens being situated in the bulge exceeds 70\%.

\section{Summary and conclusion} \label{sec:seven}
        
We analyzed the two microlensing events KMT-2021-BLG-2609 and KMT-2022-BLG-0303.  The light curves 
of these events exhibited anomalies with similar traits: short-term positive signals appearing on 
the sides of the lensing light curves of low-magnification events.  Through detailed examination 
of each event, we found that these anomalies originated from a common channel, wherein the source 
passed near the planetary caustic induced by a planet with projected separations from the host star 
exceeding the Einstein radius.

We conducted a thorough examination of the typical degeneracies often encountered in interpreting 
planetary signals arising from this channel.  Our investigation uncovered that interpreting the 
anomaly of KMT-2021-BLG-2609 was affected by the well-known "inner--outer" degeneracy. In contrast, 
despite sharing a similar lens-system configuration, interpreting the anomaly of KMT-2022-BLG-0303 
did not face this degeneracy.  This was attributed to the substantially smaller ratio of the source 
size to the caustic size for KMT-2022-BLG-0303 than that for KMT-2021-BLG-2609.  In addition to the 
inner--outer degeneracy, interpreting the anomaly in KMT-2021-BLG-2609 is subject to an additional 
degeneracy between Cannae and von Schlieffen solutions. In the Cannae solution, the source entirely 
encompasses the caustic, while in the von Schlieffen solution, the source only partially encompasses 
the caustic.  Furthermore, we found that the mass ratios of the two von Schlieffen solutions were 
systematically higher than those of the three Cannae solutions.  Noting that this was similar to 
the situation for OGLE-2017-BLG-0173 \citep{Hwang2018}, we conjectured that this is a generic 
feature of the Cannae--von Schlieffen degeneracy.

We utilized Bayesian analyses to estimate the physical parameters of the planetary systems, 
leveraging the constraints provided by the measured observables of the events.  From this analysis, 
the host of KMT-2021-BLG-2609L is determined to be a low-mass star with a mass approximately 0.2 
times that of the Sun, while the planet's mass ranges from approximately 0.032 to 0.112 times that 
of Jupiter, depending on the solutions.  For the planetary system KMT-2022-BLG-0303L, it includes 
a planet with a mass roughly half that of Jupiter and a host star with a mass approximately 0.37 
times that of the Sun.  It is notable that in both lensing events, the lenses are likely positioned 
in the bulge.

\begin{acknowledgements}
Work by C.H. was supported by the grant of National Research Foundation of Korea.
This research has made use of the KMTNet system operated by the Korea Astronomy and Space
Science Institute (KASI) at three host sites of CTIO in Chile, SAAO in South Africa, and SSO in
Australia. Data transfer from the host site to KASI was supported by the Korea Research
Environment Open NETwork (KREONET). This research was supported by KASI under the R\&D
program (Project No. 2024-1-832-01) supervised by the Ministry of Science and ICT.
W.Zang acknowledges the support from the Harvard-Smithsonian Center for Astrophysics
through the CfA Fellowship.
J.C.Y., I.G.S., and S.J.C. acknowledge support from NSF Grant No. AST-2108414. 
Y.S. acknowledges support from NSF Grant No. 2020740.
\end{acknowledgements}

\end{document}